\documentclass[aps,pre,superscriptaddress,twocolumn]{revtex4}
\usepackage[dvips,pdftex]{graphicx}
\usepackage{hyperref}
\usepackage{amsmath}

\begin{document}
\title{Collective oscillations in spatially modulated exciton-polariton condensate arrays}

\author{A.A.~Tikhomirov}
\author{O.I.~Kanakov}
\affiliation{Theory of Oscillations Department, Lobachevsky State
University of Nizhniy Novgorod, Russia}
\author{B.L.~Altshuler}
\affiliation{Department of Physics, Columbia University, New York, USA}
\author{M.V.~Ivanchenko}
\affiliation{Department of Bioinformatics, Lobachevsky State
University of Nizhniy Novgorod, Russia}

\begin{abstract}
We study collective dynamics of interacting centers of
exciton-polariton condensation in presence of spatial
inhomogeneity, as modeled by diatomic active oscillator lattices.
The mode formalism is developed and employed to derive existence
and stability criteria of plane wave solutions. It is
demonstrated that $k_0=0$ wave number mode with the binary
elementary cell on a
diatomic lattice possesses superior existence and stability
properties. Decreasing net on-site losses (balance of dissipation
and pumping) or conservative nonlinearity favors multistability of
modes, while increasing frequency mismatch between adjacent
oscillators detriments it. On the other hand, spatial
inhomogeneity may recover stability of modes at high
nonlinearities. Entering the region where all single-mode
solutions are unstable we discover subsequent transitions between
localized quasiperiodic, chaotic and global chaotic dynamics in
the mode space, as nonlinearity increases. Importantly, the last transition evokes the loss of synchronization.
These effects may determine lasing dynamics of interacting exciton-polariton
condensation centers.
\end{abstract}
\maketitle
\section{Introduction}

Recently, the phenomenon of exciton-polariton condensation in quantum wells formed by semiconductor
microcavities bordered by Bragg mirrors has been found \cite{Kasprzak, Balili}. It spurred considerable
fundamental and applied interest as a new experimental realization of
Bose-Einstein condensate and a perspective base for optical switches, miniature terahertz lasers and ultra-sensitive gyroscopes \cite{Keeling2011,Carusotto2013}.

Condensation centers (CCs) may be localized by artificially
created spatial structures or disorder \cite{Balili, Lai}. Every
CC interacts with incoherent environment, getting excited by an
incoherent pumping source and decaying due to light radiation,
also interacting with their neighbors.

Till now, most results demonstrating rich dynamics of such systems
referred to the spatially homogeneous case. Only recently Aleiner,
Altshuler, and Rubo studied a pair of nonidential CCs, interacting
through mixed Josephson and radiative coupling \cite{Aleiner}. In
particular, they determined conditions of transition to
synchronous regime that corresponds to laser generation, found
bistability regimes and the region of chaotic behavior. At the
same time, the systems containing a greater number of CC
were left beyond the focus of the study. Moreover, the implemented
approach made use of the formal analogy to equations describing a
specific spin system and cannot not be directly applied to the
systems with greater number of CC \cite{Aleiner}. Noteworthy, the studies addressing very similar equations for coupled Van-der-Pol oscillators with generic nonlinearity and coupling terms also demonstrated a rich dynamics of synchronization, multistability, and chaotic regimes \cite{Ivanchenko,Kuznetsov,Astakhov,Emelianova}. The number of oscillators, though, was still confined to a few.

In this paper we extend and study the model of interacting underdamped (weakly lasing) exciton-polariton CCs to Ginzburg-Landau-type oscillatory arrays with the spatial inhomogeneity, known as diatomic in lattice dynamics. (Linear stability analysis of plane waves in spatially homogenous
DGLE with insights into soliton and complex pattern formation has been presented in previous studies \cite{Aranson, Ravoux, Kofane2006, Kofane2008}.) Developing mode formalism for active lattices, we derive and analyze the mode excitation and instability conditions in dependence on the parameters of the system. We study transition to complex multi-mode dynamics to find the two regimes of chaos: local and global in mode space. The crossover between them appears to be related to progressing dimensionality of modulational instability of the seed mode. Transition to global chaos is accompanied by the loss of synchronization.

We generalize the model of two interacting underdamped CCs \cite{Aleiner} to a discrete Ginzburg-Landau type equation array:
\begin{equation}\label{2}
\begin{split}
\dot
z_n=&\left(-g+i\Delta(-1)^n\right)\frac{z_n}{2}-(\beta+i\alpha)|z_n|^2\frac{z_n}{2}-\\&-(\gamma-iJ)\frac{z_{n-1}+z_{n+1}}{2},
\end{split}
\end{equation}
where $z_n=\sqrt{m}e^{i\varphi}$, $m$ and $\varphi$ are occupation
and phase of the $n$-th CC respectively, $n=1,\ldots,N$, periodic
boundary conditions are assumed $z_{N+1} \equiv z_1$.
Parameter $g$ is a net local linear dissipation rate (the
difference between the escape and incoming rates of the bosons for
a CC). Throughout the paper we assume that dissipation in an
individual CC prevails, that is $g>0$, and lasing becomes possible
only due to interaction between CCs. Parameters $\alpha$ and
$\beta$ describe nonlinear frequency shift and dissipation,
$\gamma$ and $J$ are ``radiative'' \cite{Aleiner} and Josephson
coupling between adjacent CCs, respectively, all assumed to take
non-negative values. $\Delta$ is frequency mismatch between
neighbor oscillators. Without the loss
of generality we set $\beta=J=1$.

The paper is organized as follows. In section \ref{sec2} we
introduce the mode formalism and employ it to study excitation and modulational
instability of plane waves in spatially homogeneous (subsection
\ref{sec2a}) and diatomic (subsection \ref{sec2b}) arrays. In
section \ref{sec3} we study the transition from local to multi-mode chaos and demonstrate its connection to
growing number of unstable directions for the most unstable waves.

\section{Mode formalism and modulational instability}\label{sec2}
\subsection{Spatially homogeneous case}\label{sec2a}
Let us first conduct a mode analysis for the homogeneous case
$\Delta=0$. The mode variables $a_k$ are introduced by
\begin{equation} \label{3}
z_n=\frac{1}{\sqrt{N}}\sum_k a_k e^{i k n},
\end{equation}
where $k=\frac{2\pi q}{N}$ are the mode wavenumbers with integer
$q=-\frac{N-1}{2},\ldots,\frac{N-1}{2}$ for $N$ odd, or
$q=-\frac{N}{2}+1,\ldots,\frac{N}{2}$ for $N$ even. Defined this
way, the wavenumber $k$ runs a set of $N$ values in the first
Brillouin zone $k\in (-\pi,\pi]$. This transform turns (\ref{2})
into \begin{multline}\label{4} \dot a_k = -
\left[\frac{g}{2} + (\gamma-i)\cos k  \right] a_k -\\- \frac{1+ i
\alpha}{2N} \sum_{k_1 k_2 k_3} a_{k_1} a_{k_2} a_{k_3}^{\ast}
\delta_{2\pi} (k_1+k_2-k_3-k).
\end{multline}
Here the selective interaction is defined by $\delta_{2\pi}(k)=1$
if $k$ is equal to zero or a multiple of $2\pi$, and
$\delta_{2\pi}(k)=0$ otherwise. Note, that our special
statements related to the mode $k=\pi$ refer only to the case of
even $N$, when this mode exists.

These dynamical equations readily yield the excitation condition
of modes
\begin{equation}\label{5}
\frac{g}{2}+\gamma\cos k<0,
\end{equation}
which is wavenumber dependent, and is achieved, in general, with decrease of losses $g$ (or increase in pumping) or increase in radiative coupling $\gamma$.

It is easy to see that $\pi$-mode is the first one to get excited if $g<2\gamma$ and has the largest increment.
Conversely, if $g>2\gamma$ oscillations damp out and the zero equilibrium state is stable.

Noteworthy, Eq. (\ref{4}) has $N$ single-mode manifolds, namely,
solutions $a_{k_0}\neq 0,\quad a_{k\neq k_0}=0$, where $k_0$
is any permitted wavenumber. The dynamics on each manifold is described by
equation
\begin{equation}\label{6}
\dot a_{k_0} = - \left[\frac{g}{2} + (\gamma-i)\cos k_0 \right]
a_{k_0} - \frac{1+ i \alpha}{2N} |a_{k_0}|^2 a_{k_0}.
\end{equation}
If the excitation condition (\ref{5}) for that mode is
fulfilled, the dynamics on the manifold converges to a limit cycle of Eq. (\ref{6})
\begin{equation}\label{11}
a_{k_0}=A\sqrt{N}e^{i\Omega t},\quad A\in R,
\end{equation}
 where
\begin{equation}\label{7}
A^2=-g-2\gamma\cos k_0,
\end{equation}
 which is stable in the
corresponding manifold. In direct space (\ref{2}) this periodic solution
corresponds to a plane wave
\begin{equation}\label{8}
z_n(t)=Ae^{i\Omega t+ik_0n}.
\end{equation}

To investigate linear stability of these plane waves it
suffices to consider single or two side modes
perturbations, as determined by selective interaction in
(\ref{4}).

Elementary analysis of a single mode perturbations $a_k$
to the original mode $a_{k_0}$ yields a necessary stability criterion
\begin{equation}\label{9}
\frac{g}{2}+\gamma+2\gamma\cos k_0 < 0,
\end{equation}
determined again by the balance of net losses $g$, radiative
coupling $\gamma$, and is seed wave number $k_0$
specific.

Comparing (\ref{9}) to the excitation condition (\ref{5}), one
immediately recognizes that all modes except the one with the largest increment $k_0=\pi$ are
unstable on their excitation. Going away from bifurcation with
decreasing losses $g$ (equivalently, increasing pumping) or
increasing radiative coupling $\gamma$, this instability vanishes.

The exact stability criterion follows
from additionally introducing perturbations as pairs of modes,
with wave numbers symmetric relative to the wave number of the
seed wave. Linearizing Eq. (\ref{4}) in the neighborhood of a
single-mode periodic solution (\ref{11}) one obtains the increment
of a pair of modes $k=k_0+c$ and $k^{'}=k_0-c$ (see Appendix for
details):
\begin{equation}\label{10}
\begin{array}{l}
p=\frac{1}{2}Re\left[-2A^2+\lambda_k+\lambda_{k^{'}}^*+[(2A^2+\right.\\\left.-\lambda_k-\lambda_{k^{'}}^*)^2-4(i\Omega-\lambda_k+\right.\\\left.+dA^2)(-i\Omega-\lambda_{k^{'}}^*+d^*A^2)+d
d^*A^4]^{1/2}\right],
\end{array}
\end{equation}
where $\lambda_k=-\left[\frac{g}{2} +
(\gamma-i)\cos k\right]$, $d=1+i\alpha$, $\Omega=\cos k_0
-\alpha A^2/2$.

If $p<0$ for all side mode pairs, then the plane wave is linearly stable. Conversely, if $p>0$ at least for one pair of side modes the solution is unstable.

Eqs. (\ref{5}) and (\ref{10}) impose simultaneous conditions on
existence and stability of single-mode solutions, and it
is instructive for understanding to study some special cases.
First, one can explicitly resolve it for the mode $k_0=\pi$ to
derive the following condition on the strength of radiative
coupling:
\begin{equation}\label{10a}
\gamma>\alpha, \ \gamma>g/2.
\end{equation}
For $\gamma<\alpha$ and $\gamma>g/2$ the $\pi$-mode solution
exists, but is unstable. Second, assuming the balance of linear
loss and pumping $g=0$, when all modes within
$k\in[\pi/2,\pi]$ exist, according to (\ref{5}), one can determine
the boundary $k^*$ above which the modes are stable:
\begin{equation}\label{10b}
\sin^2 k^*\approx\frac{\gamma-\alpha}{\gamma-\alpha+\gamma(1+\alpha^2)},
\end{equation}
the approximation here implying $c\rightarrow0$ for the side modes. Note also an agreement with the stability condition for the $\pi$-mode (\ref{10a}), the last stable one as nonlinearity $\alpha$ increases (\ref{10b}).

\begin{figure}[ht!!!]
\begin{center}
(a)\includegraphics[width=0.9\columnwidth,keepaspectratio,clip]{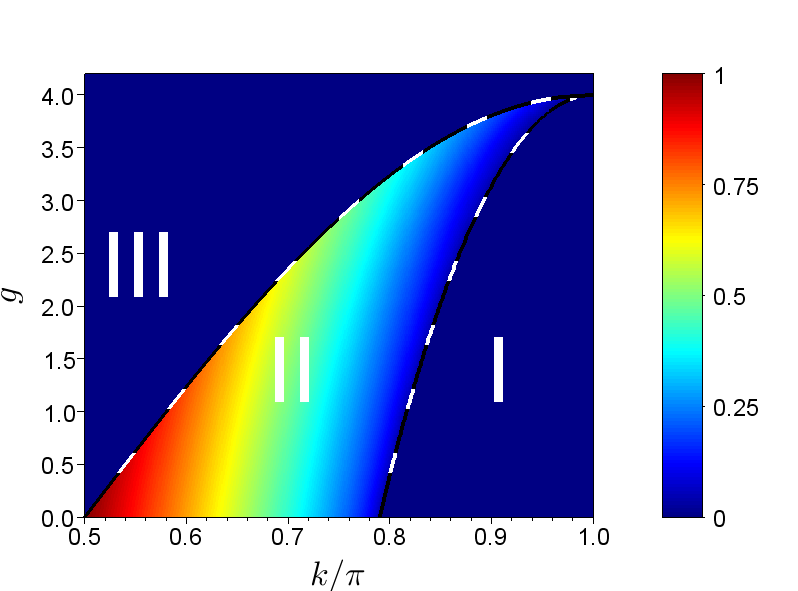}
(b)\includegraphics[width=0.9\columnwidth,keepaspectratio,clip]{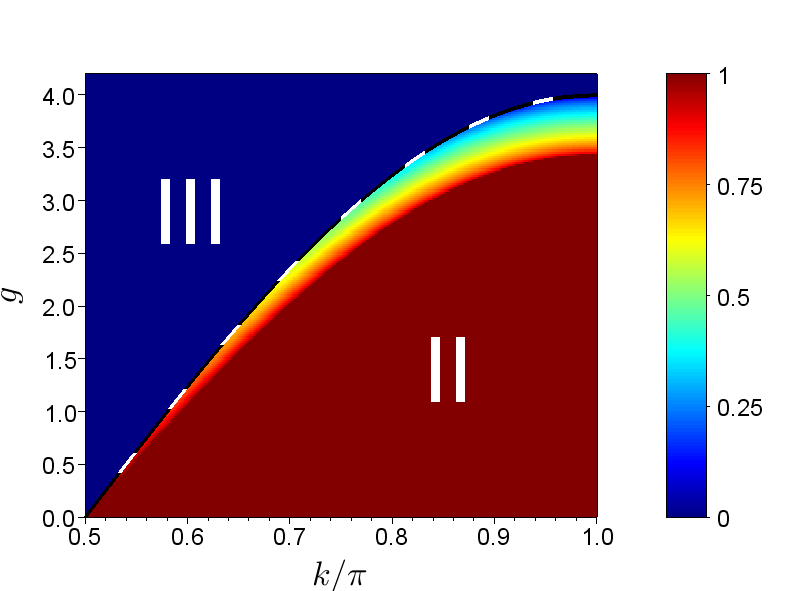}
(c)\includegraphics[width=0.9\columnwidth,keepaspectratio,clip]{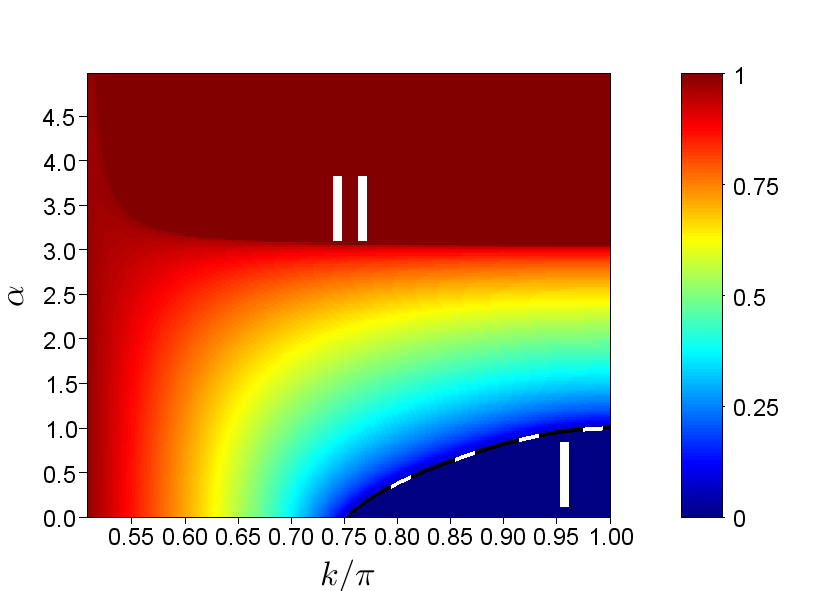}
\caption{Stability of single-mode solutions with wave numbers $k$ as the net damping rate $g$ or conservative nonlinearity $\alpha$ vary: linearly stable (I), linearly unstable (II) and absent (III). In region (II) color codes the number of unstable directions for a specific seed wave number, as given by Eq. (\ref{10}), normalized by the system size. At relatively weak nonlinearity (a) $\alpha=0.5$, $\gamma=2$ all three regions are present, while for strong nonlinearity (b) $\alpha=20$, $\gamma=2$ the region of stable single-mode solutions may not appear, in accordance with (\ref{10b}).
In general, increasing nonlinearity destabilizes single-mode solutions (c) $\gamma=1$, $g=0.05$.}\label{fig1}
\end{center}
\end{figure}

To get deeper insight into the mode dynamics, we plot diagrams in the wave number -- parameter space, distinguishing three regions, where modes with respective wave number are linearly stable (I), linearly unstable (II), and do not exist (III). The borders between these regions are obtained according to mode excitation (\ref{5}) and stability (\ref{10}) conditions. In region (II) the color codes the relative number of unstable directions for a specific seed wave number, as given by Eq. (\ref{10}).

Studying these diagrams for the net losses $g$, the simplest to control in experiment, we recover two generic cases, in accordance to Eq.(\ref{10a}), see Fig.\ref{fig1}. For relatively weak conservative nonlineary with respect to radiative coupling ($\gamma>\alpha$) all the three regions (I), (II) and (III) are present (Fig.\ref{fig1}(a)). On decreasing net losses $g$ (increasing pumping), one observes gradual birth of single-mode solutions (transition from (III) to (II)) and stabilization of some about wave number $k=\pi$ (transition from (II) to (I)). In (I) one gets high multistability of single-mode solutions, in addition to possible multi-mode attractors. Mode with $k=\pi$ is the only one to pass from (III) to (I) directly.

Conversely, relatively strong conservative nonlinearity
$\gamma<\alpha$ destabilizes single-mode solutions and only
regions (II) and (III) are observed (Fig.\ref{fig1}(b)).

Note that in the region (II) multi-mode solutions may develop,
but we leave them beyond the scope of the present study.

In general, when single-mode solutions are possible
($\gamma>g/2$), weak nonlinearity allows for a family of stable
plane waves about $k=\pi$, region (I), and multi-stability, while
increasing nonlinearity destabilizes them and evokes region (II),
see Fig.\ref{fig1}(c).

\subsection{Diatomic array}\label{sec2b}

Let us analyze existence and stability of plane wave solutions in
diatomic arrays ($\Delta\neq0$). It is convenient to introduce a
binary elementary cell: $x_n=z_{2n-1}$, $y_n=z_{2n}$ and variables
$a_k^+$,$a_k^-$
\begin{equation}\label{12}
\left\{
\begin{aligned}
x_n&=\frac{1}{\sqrt{N}}\sum\limits_{k}\left(\xi_k^+ a_k^+ +\xi_k^-
a_k^-\right)e^{ikn}\\
y_n&=\frac{1}{\sqrt{N}}\sum\limits_{k}\left(\eta_k^+ a_k^+
+\eta_k^- a_k^-\right)e^{ikn},
\end{aligned}
\right.
\end{equation}
where $k$ is the mode wave number, twice the respective wave number in the original lattice (\ref{2}):
\begin{equation}
\label{12aaa}
k_{\Delta\neq0}=2k_{\Delta=0}.
\end{equation}

\begin{figure*}[ht!!!]
\begin{center}
(a)\includegraphics[width=0.9\columnwidth,keepaspectratio,clip]{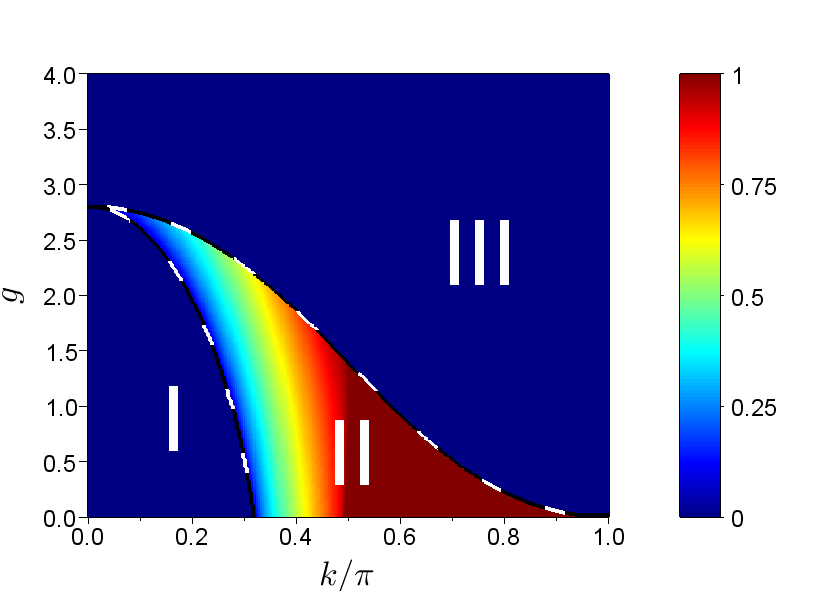}
(b)\includegraphics[width=0.9\columnwidth,keepaspectratio,clip]{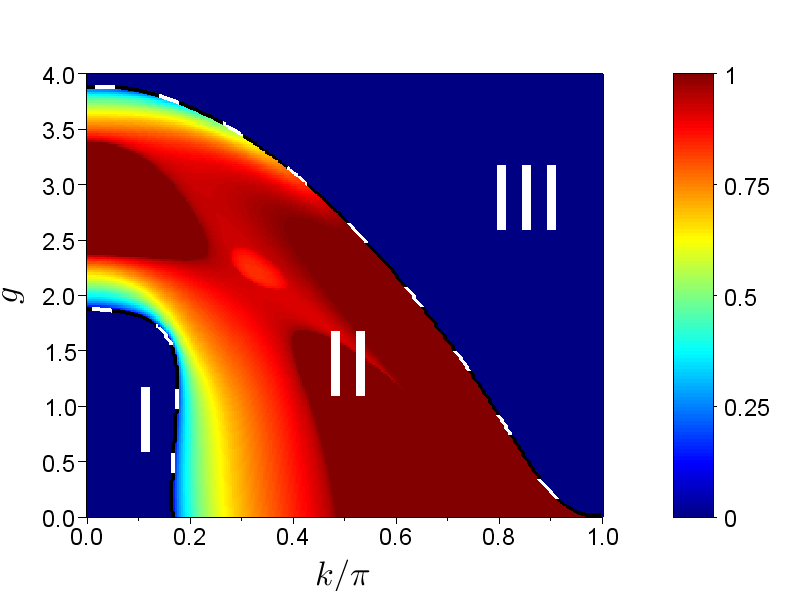}
(c)\includegraphics[width=0.9\columnwidth,keepaspectratio,clip]{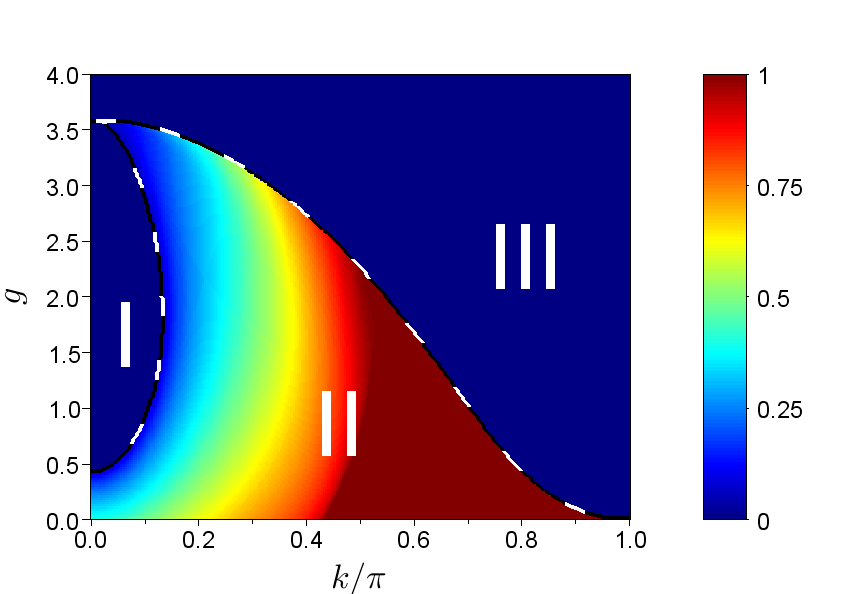}
(d)\includegraphics[width=0.9\columnwidth,keepaspectratio,clip]{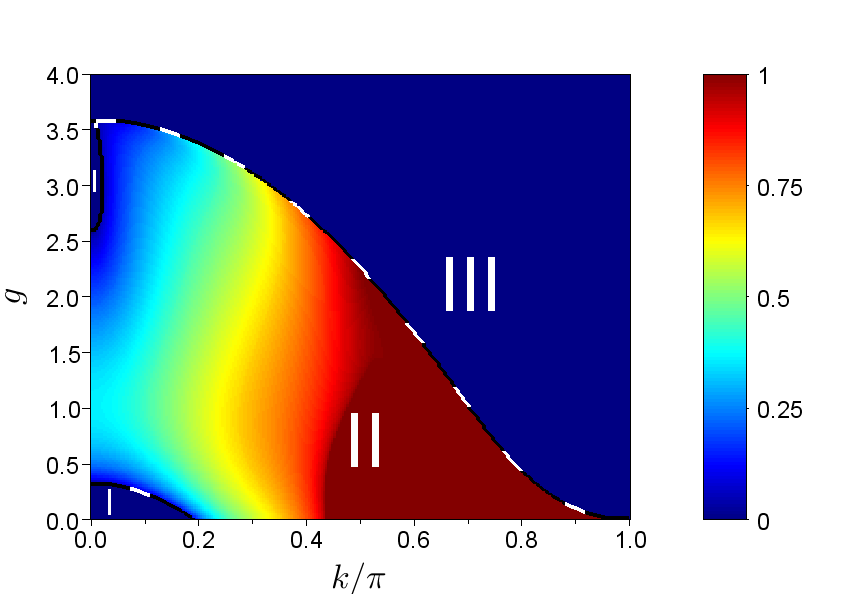}
(e)\includegraphics[width=0.9\columnwidth,keepaspectratio,clip]{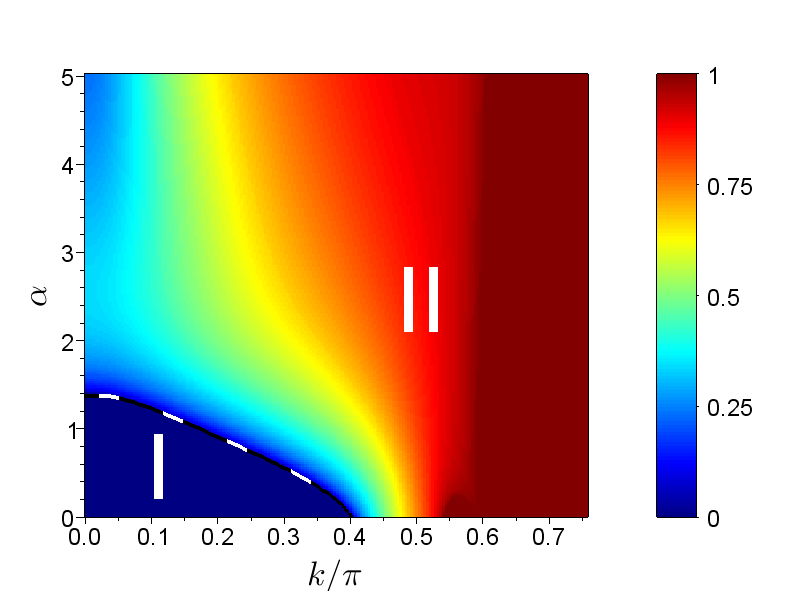}
(f)\includegraphics[width=0.9\columnwidth,keepaspectratio,clip]{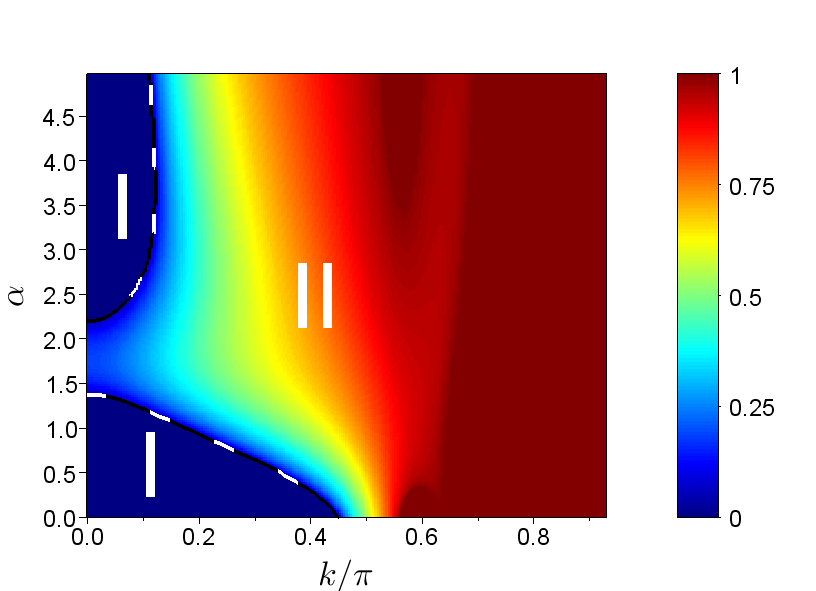}
\caption{Effects of the frequency mismatch $\Delta$ on stability and existence of single-mode solutions. Notations and color coding inherit Fig.\ref{fig1}. Keeping in mind the relation between the mode wave numbers for homogeneous and diatomic arrays (\ref{12aaa}), compare (a,b,f) to Fig.\ref{fig1}(a,b,c), respectively, and observe qualitatively new features in (c,d). Here
(a)~$\alpha=0.5$, $\Delta=3.5$, $\gamma=2$, (b)~$\alpha=20$,
$\Delta=1$, $\gamma=2$, (c)~$\alpha=1.55$, $\Delta=2$, $\gamma=2$,
(d)~$\alpha=3.1$, $\Delta=2$, $\gamma=2$, (e)~$\gamma=\Delta=1$, $g=0.5$, (f)~$\gamma=\Delta=1$, $g=0.05$.}\label{fig3}
\end{center}
\end{figure*}

Coefficients $\xi_k^{\pm}$, $\eta_k^{\pm}$ are calculated to ensure that transformation
(\ref{12}) is canonical in conservative limit ($g=\gamma=\beta=0$)
and diagonalizes the linear part of the dynamical
equations (\ref{2}):
$$\xi_k^{\pm}=\frac{\xi_k^{\pm
0}}{|\xi_k^{\pm 0}|^2+|\eta_k^{\pm 0}|^2},\quad
\eta_k^{\pm}=\frac{\eta_k^{\pm 0}}{|\xi_k^{\pm 0}|^2+|\eta_k^{\pm
0}|^2},$$
$$\xi_k^{\pm
0}=\frac{(\gamma-i)(1+e^{-ik})}{i\Delta\mp
2\sqrt{(\gamma-i)^2\cos^2\frac{k}{2}-\frac{\Delta^2}{4}}},\quad\eta_k^{\pm
0}=1,\quad k\neq\pi,$$
 $$\xi_{\pi}^{+0}=1,\eta_{\pi}^{+0}=0,\xi_{\pi}^{-0}=0,\eta_{\pi}^{-0}=1.$$

Dynamics of modes follow then
\begin{equation}\label{13}
\begin{split}
&\dot a_k^s=-i\omega_k^s
a_k^s-\frac{1+i\alpha}{2N}\times\\\times &\sum\limits_{k_1 k_2
k_3}\sum\limits_{s_1 s_2 s_3}[D_{k_1 k_2 k_3 k}^{s_1 s_2 s_3
s}a_{k_1}^{s_1}a_{k_2}^{s_2}a_{k_3}^{s_3*}\delta(k_1+k_2-k_3-k)].
\end{split}
\end{equation}

Here $s_i$ denotes the signature ``$+$'' or ``$-$'', and
coefficients $D_{k_1 k_2 k_3 k}^{s_1 s_2 s_3 s}$ determine
selective interaction between the modes:
$$D_{k_1 k_2 k_3 k}^{s_1 s_2 s_3
+}=\frac{1}{\xi_k^-\eta_k^+
-\eta_k^-\xi_k^+}(\xi_k^-\eta_{k_1}^{s_1}\eta_{k_2}^{s_2}\eta_{k_3}^{s_3*}-\eta_k^-\xi_{k_1}^{s_1}\xi_{k_2}^{s_2}\xi_{k_3}^{s_3*}),$$
$$D_{k_1 k_2 k_3 k}^{s_1 s_2 s_3 -}=\frac{1}{\xi_k^-\eta_k^+
-\eta_k^-\xi_k^+}(\eta_k^+\xi_{k_1}^{s_1}\xi_{k_2}^{s_2}\xi_{k_3}^{s_3*}-\xi_k^+\eta_{k_1}^{s_1}\eta_{k_2}^{s_2}\eta_{k_3}^{s_3*}),$$

Frequencies $\omega_k^s$ follow dispersion equation
\begin{equation}\label{12a}
\left(-i\omega_k^{\pm}+\frac{g}{2}\right)^2=(\gamma-i)^2\cos^2\frac{k}{2}-\frac{\Delta^2}{4}.
\end{equation}

Eq. (\ref{13}) gives excitation condition for modes $a_k^s$, simultaneously met for both signatures $s$:
\begin{equation}\label{14}
\mbox{Im}(\omega_k^s)>0
\end{equation}
It demonstrates that diatomic system possesses $N$ two-mode manifolds, i.e.
 solutions of the form $a_{k_0}^{\pm}\neq 0$, $a_{k\neq
k_0}^{\pm}=0$, rather than single-mode ones as in the homogeneous case. The only exceptions
correspond to $a_{\pi}^+$ and $a_{\pi}^-$ modes.

Excitation threshold can be found letting $\mbox{Im}(\omega_k^s)=0$, which after some algebra gives
\begin{equation}\label{14a}
(g/2)^4+\left[\frac{\Delta^2}{4}+(1-\gamma^2)\cos^2\frac{k}{2}\right](g/2)^2-\gamma^2 \cos^4\frac{k}{2}=0,
\end{equation}
and several corollaries follow. First, for all $\gamma\neq0,
k\neq\pi$ there exists a unique threshold $g=g^*(k)$,
below which the zero state becomes unstable with respect to
$a_{k_0}^{\pm}$. Moreover, one finds that this threshold is
maximized by $k=0$, in agreement to the homogeneous limit
$\Delta=0$, where the $\pi$-mode is the first to get excited
(recall that we currently work with the binary elementary cell
 and wave numbers double (\ref{12aaa})). Addressing the
effect of the frequency mismatch, one readily recognizes that the
excitation threshold $g=g^*(k,\Delta)$ is a decreasing function of
$\Delta$, as given by (\ref{14a}). It means that non-zero spatial
inhomogeneity requires to reduce net losses in order to obtain
excitation of modes. Notably, in absence of dissipative coupling
$\gamma=0$ the zero solution will always be stable.

Once excitation condition (\ref{14}) for a particular wave number $k_0$ is attained, there emerges a periodic trajectory in the corresponding two-mode manifold:
\begin{equation}\label{15}
a_{k_0}^{\pm}=\sqrt{N}A_{1,2}e^{-i\Omega t},
\end{equation}
whose amplitudes $A_{1,2}$ get saturated by nonlinear dissipation
in (\ref{13}) and can be numerically calculated (see Appendix). In
the original equations (\ref{2}) one recovers a plane wave
solution
\begin{equation}\label{16} \left\{
\begin{aligned}
x_n=B_1 e^{ik_0 n-i\Omega t}\\
y_n=B_2 e^{ik_0 n-i\Omega t}
\end{aligned}
\right.
\end{equation}
where $B_1=\xi_k^+A_1+\xi_k^-A_2$, $B_2=\eta_k^+A_1+\eta_k^-A_2$.

As before, we study modulational instability of these waves against arbitrary small perturbations in the side modes $a_k^{\pm}$, $k\neq k_0$, and $k_0$ referring to the seed mode. Evolution of perturbations follows from
linearization of Eq. (\ref{13}) in neighborhood of a periodic solution (\ref{15}). It is represented by mutually independent sets of dynamical equations for quadruplets of side modes $a_k^{\pm}$, $a_{k^{'}}^{\pm}$, where
$k=k_0+c$, $k^{'}=k_0-c$ (see Appendix for details).

The seed mode solution of (\ref{13}) is linearly stable if all increments of the side modes are negative and, additionally, stability within the two-mode manifold (\ref{15}) is provided. If an increment in at least one side modes quadruplet is positive or instability within the manifold develops, the seed solution becomes unstable.

Since analytical studies in the general case are extremely complicated, we further resort to a computationally exact numerical analysis (numerical calculation of the amplitudes of pure mode solutions and diagonalization of stability matrices about them).
The resulting typical stability diagrams demonstrate important qualitative distinctions from the spatially homogeneous case, which are illustrated in
Fig.~\ref{fig3}.

First, spatial inhomogeneity decreases the regions, where mode solutions exist, stable (I) or unstable (II), in agreement to (\ref{14a}), compare the thresholds in $g$ for each $k$ in  Fig.\ref{fig1}(a) and Fig.\ref{fig3}(a). At the same time, inhomogeneity may improve stability of these solutions at relatively high nonlinearities, leading to emergence of the region (I) within (II), compare Fig.\ref{fig1}(b) and Fig.\ref{fig3}(b). Two principally new features may appear in $(k,g)$ diagrams. For $\gamma>\alpha$, when all the three regions (I)-(III) are present for $\Delta=0$, inhomogeneity can decrease the stability region (I), making it completely detached from $g=0$ axis (Fig.\ref{fig3}(c)). For $\gamma<\alpha$ the stability region (I) may emerge in two disconnected areas within (II), as demonstrated in Fig.\ref{fig3}(d). Turning to $(k,\alpha)$ diagrams, one observes pictures either qualitatively similar to $\Delta=0$ (compare Fig.\ref{fig1}(c) and Fig.\ref{fig3}(e)), or a stabilizing effect of inhomogeneity, leading to emergence of the second stability island (I) within (II) (Fig.\ref{fig3}(e)).

\section{Instability and mode dynamics: from local to global chaos}\label{sec3}

Having obtained stability conditions for plane wave
solutions, we turn to the dynamics, when instability occurs. While
there appears to be a variety of complex dynamical regimes, we
focus on transition from a few to multi-mode oscillations related
to the development of dynamical chaos. This crossover presents a
special interest due to the physically relevant interpretation of
emerging broad spectrum lasing of exciton-polariton condensates.

Evolution from arbitrary initial conditions depends on attractors
of the system and their basin of attraction. If plane
waves about $k=0$ in the diatomic array are linearly stable
(region (I) in Fig.\ref{fig3}), then initial conditions slightly
perturbed from such a mode will converge to it. In this region,
therefore, multistability takes place. As the parameters vary, the
stability interval in wave numbers may decrease until the most
stable mode $k=0$ undergoes modulational instability.

\begin{figure*}[th!!!]
\begin{center}
(a)\includegraphics[width=0.9\columnwidth,keepaspectratio,clip]{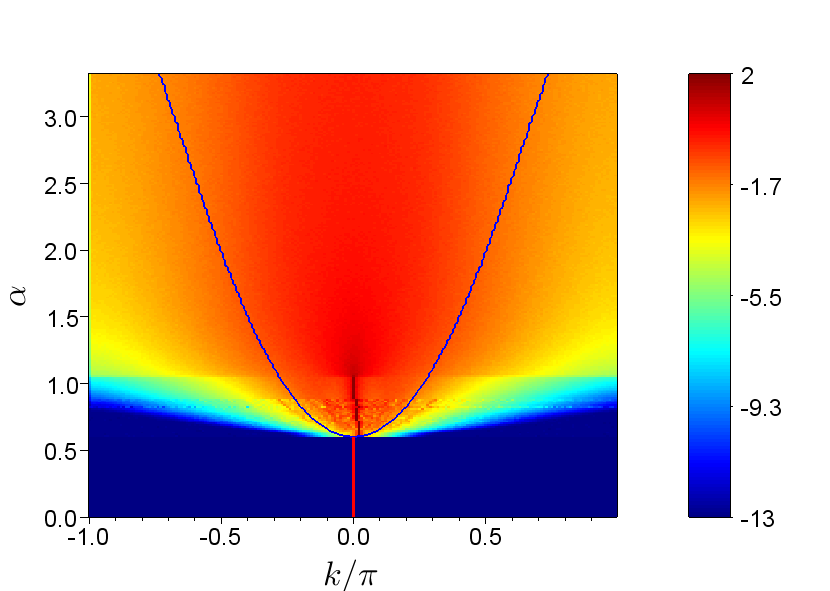}
(b)\includegraphics[width=0.9\columnwidth,keepaspectratio,clip]{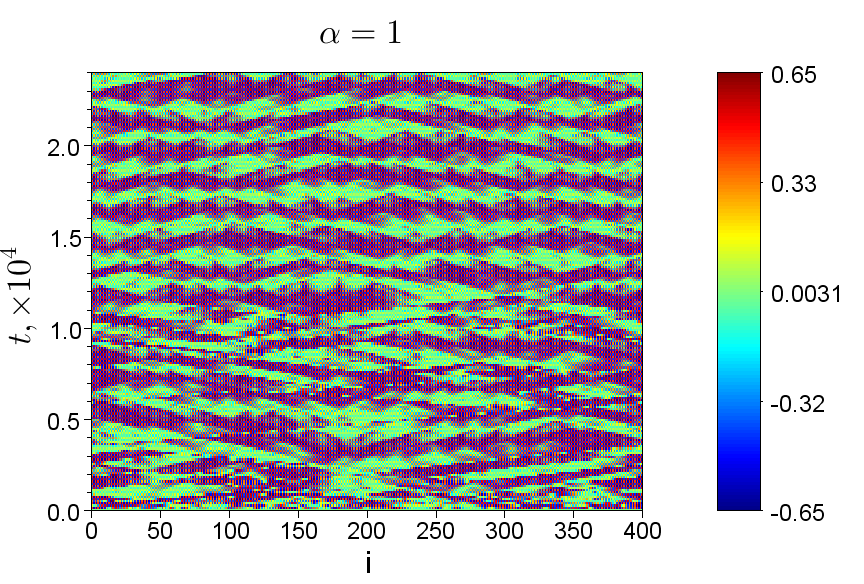}
(c)\includegraphics[width=0.9\columnwidth,keepaspectratio,clip]{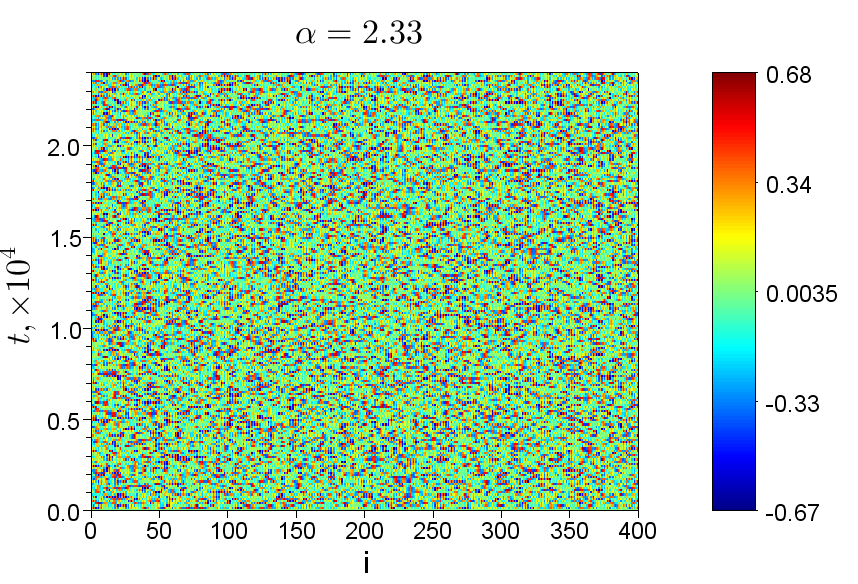}
(d)\includegraphics[width=0.9\columnwidth,keepaspectratio,clip]{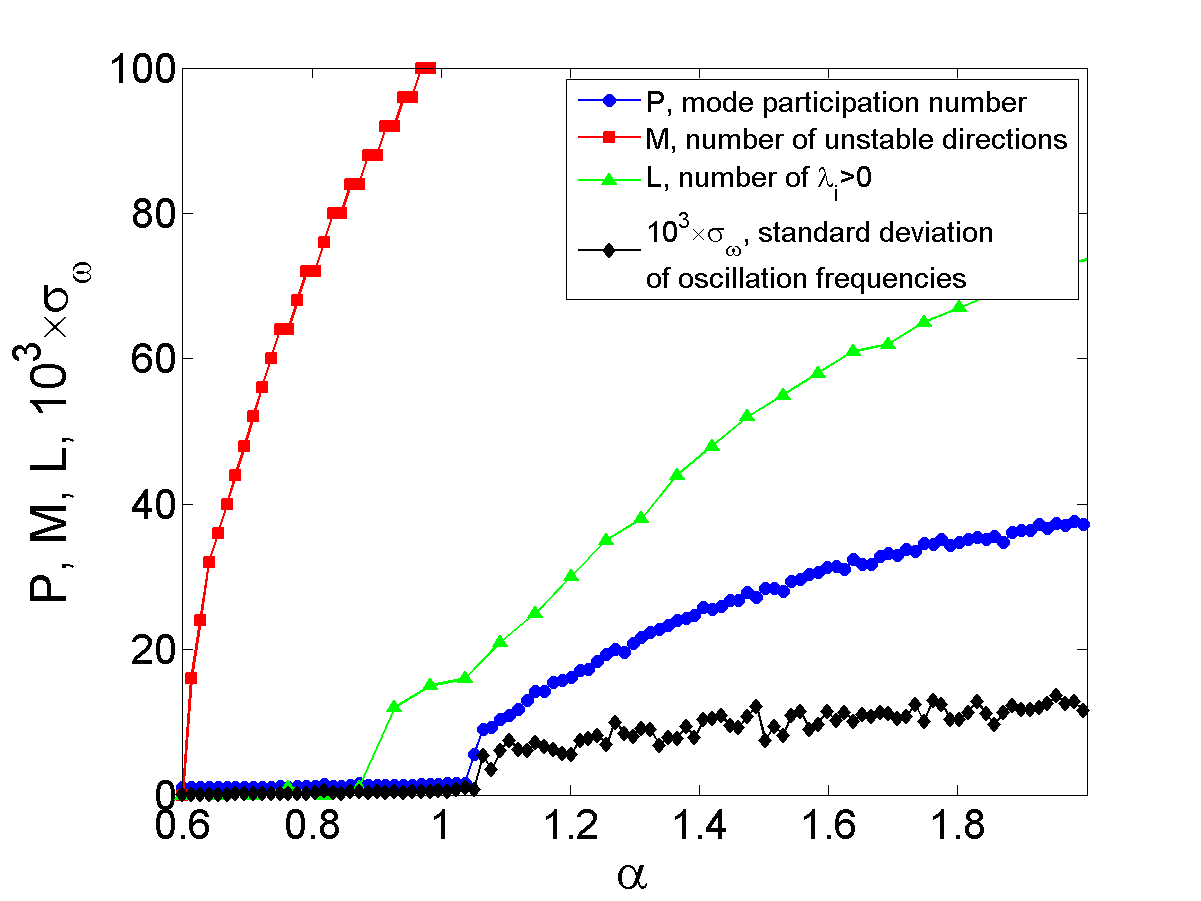}
\caption{From stable plane waves to
multi-mode chaos: (a) color coded distribution of mode space
excitations $\ln|a_k^+|^2$ ($|a_k^+|$ time-averaged after a
transient time) as nonlinearity $\alpha$ increases and
modulational instability develops, solid lines mark modulational
instability domain of the mode $k_0=0$; (b) and (c) color coded
spatio-temporal dynamics of $\mbox{Re}(z_n)$ for $\alpha=1$
(somewhat above instability threshold) and $\alpha=2.667$ (deep
into the instability region), respectively; (d) time-averaged mode
participation number $P$, the number of unstable directions $M$ of the mode $k_0=0$, the number
of positive Lyapunov exponents $L$, and the standard deviation of average oscillator frequencies versus nonlinearity $\alpha$ (note its upscaling). Here
$\gamma=0.5$, $g=0.0667$, $\Delta=0.667$, the
system size equals $N=400$. }\label{fig4}
\end{center}
\end{figure*}

To study this transition one needs to characterize the distribution of energy density in mode space. For that it is convenient to calculate participation number, the quantity defined as
\begin{equation}
P=\frac{\left(\sum_k (|a_k^+|^2+|a_k^-|^2)\right)^2}{\sum_k
(|a_k^+|^4+|a_k^-|^4)}
\end{equation}
 and estimating the number of effectively excited modes.

Fig.\ref{fig4} demonstrates a characteristic evolution of the mode participation number on
an attractor as it evolves from the stable $k=0$ mode into the instability region, while the on-site nonlinearity $\alpha$ increases.
There we observe that the excitation in the mode space closely follows modulational instability boundary for the side modes (Fig.\ref{fig4}(a)).
In direct space collective oscillations demonstrate the crossover from distorted plain waves (somewhat above instability threshold, Fig.\ref{fig4}(b)) to strongly chaotic patterns (deep in the instability region, Fig.\ref{fig4}(c)). Remarkably, the transition does not appear entirely smooth: in the interval $\alpha\approx 0.6\ldots 1.1$ the excitation is concentrated about the seed mode $k_0=0$, while at $\alpha\approx1.1$ quite a sharp break up towards a broad excitation across the whole instability region is detected.

To quantify the relation between instability and multi-mode dynamics we plot corresponding mode participation number $P$, the number of side modes $M$ with positive increment, the number $L$ of positive Lyapunov exponents $\lambda_i>0$, and standard deviation of average oscillator frequencies $\sigma_\omega$ against nonlinearity strength $\alpha$ in Fig.\ref{fig4}(d). The results suggest that modulational instability of $k_0=0$ mode leads first to quasiperiodic oscillations on a torus of a low effective dimension, evidenced by small participation number in the mode space $P\approx1$ and absence of reliably positive Lyapunov exponents in $0.6<\alpha<0.87$. Then one observes transition to dynamical chaos that may be characterized as ``local'' in the mode space in $0.97<\alpha<1.04$, since $P\approx1$ remains. In direct space this is seen as chaotic modulations of a plain wave solution (Fig.\ref{fig4}(b)). Finally, the transition to ``global'' chaos occurs at $\alpha\approx1.04$, manifested by an abrupt increase of participation number to $P\approx10$. Simultaneously, synchronization of oscillation frequencies in the direct space is broken up. Further, dimensionality of chaos and the number of effectively excited modes grow in parallel (Fig.\ref{fig4}(d)), oscillations in the direct space become irregular (Fig.\ref{fig4}(c)).

\section{Conclusion}
We have studied collective dynamics in oscillatory arrays with
diatomic spatial inhomogeneity, radiative and Josephson coupling,
conservative and dissipative nonlinearities. We developed the mode
analysis for such  systems to derive mode excitation and
modulation instability conditions. It was shown that the most
persistent and stable mode has the wave number $k_0=0$ in the
binary node basis (corresponding to the $k_0=\pi$ mode in the
limit of a homogeneous array in a singe-node basis). Generically,
the number of stable modes in vicinity of $k_0=0$ increases as the
net local damping (balance of pumping and dissipation) decreases,
giving rise to high-order multistability. Increasing conservative
nonlinearity supports the development of modulation instability of
the modes. The frequency mismatch between neighbor sites imposes
more stringent conditions on mode existence, as compared to the
homogeneous array. At the same time, inhomogeneity can
stabilize the modes at high nonlinearity,
which in the homogeneous case would be unstable. We also
identified the cases when the region of stable mode solutions
consists of two disjoint islands.

We explored the dynamics beyond the mode stability region and demonstrated several crossovers as the nonlinearity and degree of seed mode instability grow. First, the dynamics remains regular and quasiperiodic, localized in the mode space. Second, the transition to low-dimensional chaos, also localized in the mode space, is observed. Above the last threshold the mode participation number increases abruptly, oscillations span over most of the modulation instability wave interval, producing multi-dimensional chaos and the break up of synchronization.

Beside their fundamental interest for nonlinear dynamics, these
results are applicable to understanding and designing
collective behavior of active structured physical media with
spatial inhomogeneity, such as exciton-polaritons condensation
arrays, our primary system of interest, or actively coupled
waveguides.

A.T., O.K., and M.I. acknowledge support of Russian Foundation for Basic Research grant No. 13-02-97028. M.I. also acknowledges support of Russian Ministry of Science and Education (research of Section III supported by Research Assignment No. 1.115.2014/K). We thank Yu. Rubo, S. Flach, and N. Berloff for insightful discussions.

\appendix*
\section{Modulational instability
analysis} In case of spatially homogeneous system, linearization
of Eq. (\ref{4}) in the neighborhood of the mode $a_{k_0}$
(\ref{11}) leads to dynamical equations for perturbations in side
mode pairs $a_k, a_{k^{'}}$ ($k=k_0+c$, $k^{'}=k_0-c$):
\begin{equation}\label{a1} \left\{
\begin{aligned}
&\dot a_k = (\lambda_k - d A^2)a_k - \frac{d}{2}A^2 e^{2i\Omega
t}a_{k^{'}}^*,
 \\ &\dot a_{k^{'}}^* =
(\lambda_{k^{'}}^* -d^* A^2)a_{k^{'}}^* - \frac{d^*}{2} A^2
e^{-2i\Omega t}a_k,
\end{aligned}
\right.
\end{equation}
where
$\lambda_k=-\left[\frac{g}{2} + (\gamma-i)\cos k\right]$,
$d=1+i\alpha$. Substituting $a_k=y_1e^{i\Omega t}$,
$a_{k^{'}}^*=y_2e^{-i\Omega t}$ in (\ref{a1}) one obtains the
linear system of differential equations with constant
coefficients for $y_{1,2}$, whose maximal eigenvalue
yields the increment of modes $a_k,a_{k^{'}}$ (\ref{10}).

For the diatomic system from Eq. (\ref{13}) we obtain equations of
motion in the two-mode manifold with wavenumber $k_0$
\begin{widetext}
\begin{equation}\label{a2}
\begin{split}
\dot a_{k_0}^{+} =
-i\omega_{k_0}^{+}a_{k_0}^{+}-\frac{i\alpha+\beta}{2N}(D_{k_0 k_0
k_0 k_0}^{++++}|a_{k_0}^{+}|^2 a_{k_0}^{+}+2D_{k_0 k_0 k_0
k_0}^{-+++}|a_{k_0}^{+}|^2a_{k_0}^{-}+D_{k_0 k_0 k_0
k_0}^{++-+}a_{k_0}^{{+}^2} a_{k_0}^{-*}+\\+2D_{k_0 k_0 k_0
k_0}^{+--+}|a_{k_0}^{-}|^2 a_{k_0}^{+}+D_{k_0 k_0 k_0 k_0}^{--++}
a_{k_0}^{{-}^2}a_{k_0}^{+*}+D_{k_0
k_0 k_0 k_0}^{---+}|a_{k_0}^{-}|^2 a_{k_0}^{-})\\
\dot a_{k_0}^{-} =
-i\omega_{k_0}^{-}a_{k_0}^{-}-\frac{i\alpha+\beta}{2N}(D_{k_0 k_0
k_0 k_0}^{+++-}|a_{k_0}^{+}|^2 a_{k_0}^{+}+2D_{k_0 k_0 k_0
k_0}^{-++-}|a_{k_0}^{+}|^2a_{k_0}^{-}+D_{k_0 k_0 k_0
k_0}^{++--}a_{k_0}^{{+}^2} a_{k_0}^{-*}+\\+2D_{k_0 k_0 k_0
k_0}^{+---}|a_{k_0}^{-}|^2 a_{k_0}^{+}+D_{k_0 k_0 k_0 k_0}^{--+-}
a_{k_0}^{{-}^2}a_{k_0}^{+*}+D_{k_0 k_0 k_0
k_0}^{----}|a_{k_0}^{-}|^2 a_{k_0}^{-}).
\end{split}
\end{equation}
\end{widetext}
If periodic solution (\ref{15}) exists in this two-mode manifold
we may find values $A_{1,2}$ by substituting it into Eq.
(\ref{a2}).

Linearization of Eq. (\ref{13}) in neighborhood of (\ref{15})
gives a set of self-contained systems for quadruplets of side
modes $a_k^{\pm}$, $a_{k^{'}}^{\pm}$ where $k=k_0+c$,
$k^{'}=k_0-c$. Dynamics of a quadruplet is governed by
\begin{widetext}
\begin{equation}\label{a3}
\begin{split}
&\dot a_k^+=-i\omega_k^+a_k^+-\frac{i\alpha+\beta}{2}\left[2D_{k
k_0 k_0 k}^{++++}|A_1|^2a_k^++2D_{k k_0 k_0
k}^{-+++}|A_1|^2a_k^-+D_{k_0 k_0 k^{'} k}^{++++}A_1^2e^{-2i\Omega
t}a_{k^{'}}^{+*}+D_{k_0 k_0 k^{'} k}^{++-+}A_1^2e^{-2i\Omega
t}a_{k^{'}}^{-*}+\right.\\&+2A_2\left(D_{k k_0 k_0
k}^{+-++}A_1^*a_k^++D_{k k_0 k_0 k}^{--++}A_1^*a_k^-+D_{k k_0 k_0
k}^{++-+}A_1a_k^++D_{k k_0 k_0 k}^{-+-+}A_1a_k^-+D_{k_0 k_0 k^{'}
k}^{+-++}A_1e^{-2i\Omega t}a_{k^{'}}^{+*}+\right.\\&
\left.\left.+D_{k_0 k_0 k^{'} k}^{+--+}A_1e^{-2i\Omega
t}a_{k^{'}}^{-*}\right)+A_2^2\left(2D_{k k_0 k_0
k}^{+--+}a_k^++2D_{k k_0 k_0 k}^{---+}a_k^-+D_{k_0 k_0 k^{'}
k}^{--++}e^{-2i\Omega t}a_{k^{'}}^{+*}+D_{k_0 k_0 k^{'}
k}^{---+}e^{-2i\Omega t}a_{k^{'}}^{-*}\right)\right]\\& \dot
a_k^-=-i\omega_k^-a_k^--\frac{i\alpha+\beta}{2}\left[2D_{k k_0 k_0
k}^{+++-}|A_1|^2a_k^++2D_{k k_0 k_0 k}^{-++-}|A_1|^2a_k^-+D_{k_0
k_0 k^{'} k}^{+++-}A_1^2e^{-2i\Omega t}a_{k^{'}}^{+*}+D_{k_0 k_0
k^{'} k}^{++--}A_1^2e^{-2i\Omega
t}a_{k^{'}}^{-*}+\right.\\&+2A_2\left(D_{k k_0 k_0
k}^{+-+-}A_1^*a_k^++D_{k k_0 k_0 k}^{--+-}A_1^*a_k^-+D_{k k_0 k_0
k}^{++--}A_1a_k^++D_{k k_0 k_0 k}^{-+--}A_1a_k^-+D_{k_0 k_0 k^{'}
k}^{+-+-}A_1e^{-2i\Omega t}a_{k^{'}}^{+*}+\right.\\&
\left.\left.+D_{k_0 k_0 k^{'} k}^{+---}A_1e^{-2i\Omega
t}a_{k^{'}}^{-*}\right)+A_2^2\left(2D_{k k_0 k_0
k}^{+---}a_k^++2D_{k k_0 k_0 k}^{----}a_k^-+D_{k_0 k_0 k^{'}
k}^{--+-}e^{-2i\Omega t}a_{k^{'}}^{+*}+D_{k_0 k_0 k^{'}
k}^{----}e^{-2i\Omega t}a_{k^{'}}^{-*}\right)\right]\\& \dot
a_{k^{'}}^{+*}=i\omega_{k^{'}}^{+*}a_{k^{'}}^{+*}+\frac{i\alpha-\beta}{2}\left[2(D_{k^{'}
k_0 k_0 k^{'}}^{++++})^*|A_1|^2a_{k^{'}}^{+*}+2(D_{k^{'} k_0 k_0
k^{'}}^{-+++})^*|A_1|^2a_{k^{'}}^{-*}+(D_{k_0 k_0 k
k^{'}}^{++++})^*A_1^{*2}e^{2i\Omega t}a_{k}^++\right.\\&+(D_{k_0
k_0 k k^{'}}^{++-+})^*A_1^{*2}e^{2i\Omega
t}a_{k}^-+2A_2\left((D_{k^{'} k_0 k_0
k^{'}}^{+-++})^*A_1a_{k^{'}}^{+*}+(D_{k^{'} k_0 k_0
k^{'}}^{--++})^*A_1a_{k^{'}}^{-*}+(D_{k^{'} k_0 k_0
k^{'}}^{++-+})^*A_1^*a_{k^{'}}^{+*}+\right.\\&\left.+(D_{k^{'} k_0
k_0 k^{'}}^{-+-+})^*A_1^*a_{k^{'}}^{-*}+(D_{k_0 k_0 k
k^{'}}^{+-++})^*A_1^*e^{2i\Omega t}a_{k}^++(D_{k_0 k_0 k
k^{'}}^{+--+})^*A_1^*e^{2i\Omega
t}a_{k}^-\right)+A_2^2\left(2(D_{k^{'} k_0 k_0
k^{'}}^{+--+})^*a_{k^{'}}^{+*}+\right.\\&\left.\left.+2(D_{k^{'}
k_0 k_0 k^{'}}^{---+})^*a_{k^{'}}^{-*}+(D_{k_0 k_0 k
k^{'}}^{--++})^*e^{2i\Omega t}a_{k}^++(D_{k_0 k_0 k
k^{'}}^{---+})^*e^{2i\Omega t}a_{k}^-\right)\right]\\& \dot
a_{k^{'}}^{-*}=i\omega_{k^{'}}^{-*}a_{k^{'}}^{-*}+\frac{i\alpha-\beta}{2}\left[2(D_{k^{'}
k_0 k_0 k^{'}}^{+++-})^*|A_1|^2a_{k^{'}}^{+*}+2(D_{k^{'} k_0 k_0
k^{'}}^{-++-})^*|A_1|^2a_{k^{'}}^{-*}+(D_{k_0 k_0 k
k^{'}}^{+++-})^*A_1^{*2}e^{2i\Omega t}a_{k}^++\right.\\&+(D_{k_0
k_0 k k^{'}}^{++--})^*A_1^{*2}e^{2i\Omega
t}a_{k}^-+2A_2\left((D_{k^{'} k_0 k_0
k^{'}}^{+-+-})^*A_1a_{k^{'}}^{+*}+(D_{k^{'} k_0 k_0
k^{'}}^{--+-})^*A_1a_{k^{'}}^{-*}+(D_{k^{'} k_0 k_0
k^{'}}^{++--})^*A_1^*a_{k^{'}}^{+*}+\right.\\&\left.+(D_{k^{'} k_0
k_0 k^{'}}^{-+--})^*A_1^*a_{k^{'}}^{-*}+(D_{k_0 k_0 k
k^{'}}^{+-+-})^*A_1^*e^{2i\Omega t}a_{k}^++(D_{k_0 k_0 k
k^{'}}^{+---})^*A_1^*e^{2i\Omega
t}a_k^-\right)+A_2^2\left(2(D_{k^{'} k_0 k_0
k^{'}}^{+---})^*a_{k^{'}}^{+*}+\right.\\&\left.\left.+2(D_{k^{'}
k_0 k_0 k^{'}}^{----})^*a_{k^{'}}^{-*}+(D_{k_0 k_0 k
k^{'}}^{--+-})^*e^{2i\Omega t}a_k^++(D_{k_0 k_0 k
k^{'}}^{----})^*e^{2i\Omega t}a_k^-\right)\right]
\end{split}
\end{equation}
\end{widetext}
Without the loss of generality we assume that $A_1$ is complex and
$A_2$ is real. Substituting $a_k^+=u_1e^{-i\Omega t}$,
$a_k^-=u_2e^{-i\Omega t}$, $a_{k^{'}}^{+*}=u_3e^{i\Omega t}$,
$a_{k^{'}}^{-*}=u_4e^{i\Omega t}$ we obtain the linear system of differential
equations with constant coefficients for
$u_{1,2,3,4}$, the maximal eigenvalue of which being the increment
of the modes $a_k^{\pm},a_{k^{'}}^{\pm}$. It is calculated
numerically.

\bibliography{literature}

\end{document}